\documentstyle[aps]{revtex}
\input epsf
\input psfig
\begin{document}


\draft
\title{Constraining Inflation with Cosmic Microwave Background Polarization}
\author{William H.\ Kinney\thanks{Electronic address: {\tt kinneyw@fnal.gov}}}
\address{NASA/Fermilab Astrophysics Center}
\address{Fermi National Accelerator Labaratory, Batavia, IL 60510}
\author{FERMILAB--PUB--98/187--A, astro-ph/9806259}
\date{June 18, 1998}
\maketitle

\begin{abstract}
Observations of the polarization of the cosmic microwave backround (CMB) have the potential to place much tighter constraints on cosmological parameters than observations of the fluctuations in temperature alone. We discuss using CMB polarization to constrain parameters relevant for distinguishing among popular models for cosmological inflation, using the MAP and Planck satellite missions as example cases. Of particular interest is the ability to detect tiny contributions to the CMB anisotropy from tensor modes, which is fundamentally limited by cosmic variance in temperature-only observations. The ability to detect a tensor/scalar ratio $r \sim 0.01$ would allow precision tests of interesting inflation models, and is possible with a modest increase in sensitivity over that planned for the Planck satellite, or potentially by ground-based experiments.
\end{abstract}

\pacs{98.80.Cq,98.70.Vc,98.80.Es}

\section{Introduction}
\label{secintro}

The Cosmic Microwave Background (CMB) radiation promises to be a powerful tool in understanding cosmology. The exact form of the anisotropy in the CMB depends on a host of cosmological parameters, such as the density $\Omega_0$, Hubble constant $H_0$, baryon density $\Omega_{\rm B}$, and so forth. A great deal has been written about using the CMB to constrain cosmological parameters \cite{bond94,knox94,knox95,jungman96,bond97,zaldarriaga97a,eisenstein98}. This paper is an extension of earlier work \cite{dodelson97}, which discussed CMB parameter estimation from the point of view of parameters relevant for distinguishing among models of cosmological inflation. Inflation \cite{guth81} has become {\it the} paradigm for understanding the initial conditions for structure formation and for CMB anisotropy. In the inflationary picture, primordial density and gravity-wave fluctuations are created from quantum fluctuations ``redshifted'' out of the horizon during an early period of superluminal expansion of the universe. The density (or {\it scalar}) fluctuations form the seeds for structure formation, and along with the gravity-wave (or {\it tensor}) fluctuations are also responsible for the observed temperature fluctuations in the CMB. Most (but not all) inflation models predict a geometrically flat universe and a nearly scale-invariant spectrum of density perturbations. Within this general framework, different inflation models make different predictions for the amplitudes and spectral indices of tensor and scalar fluctuations. These parameters can be constrained through observation of CMB temperature anisotropy, and it will be possible with upcoming CMB measurements to falsify models of inflation.

Here we consider CMB polarization as well as temperature anisotropy as a tool for constraining inflation, using NASA's MAP satellite \cite{MAPurl} and the ESA's Planck Surveyor \cite{Planckurl} as examples of the accuracy that will be achievable in the next few years. Polarization of the CMB is a generic prediction of any model in which the CMB is created by primordial density fluctuations, so that an experiment of sufficient sensitivity will almost certainly detect it. (For a pedagogical review of CMB polarization, see Ref. \cite{hu97a}. More formal treatments can be found in Refs. \cite{kosowsky96,zaldarriaga97b,kamionkowski96,hu97b}. Refs. \cite{spergel97,kamionkowski97,copeland97} discuss issues specifically related to inflation.) Observation of CMB polarization has the potential to provide us with a great deal more information than can be obtained from observation of temperature alone, and therefore has the potential to greatly strengthen constraints on parameters. The purpose of this paper is to examine quantitatively the improvements in parameter estimation that can be gained through observation of CMB polarization, with emphasis on parameters relevant for distinguishing among the ``zoo'' of currently popular inflation models. We come to two main conclusions.  First, cosmic variance in temperature-only measurements severely limits the ability to detect a small tensor/scalar ratio, and measurement of polarization allows significant improvement in the ability to study models that predict very small tensor contributions to the CMB. Second, reionization of the universe can significantly degrade the sensitivity of parameter estimation if temperature anisotropy alone is measured, but measurement of polarization effectively removes the parameter degeneracy between tensor amplitude and reionization optical depth.

The structure of this paper is as follows: Section \ref{secinflationreview} contains a brief review of cosmological inflation and the generation of perturbations. Section \ref{seccmbreview} discusses fluctuations in the CMB and how to quantify the expected measurement errors for planned experiments. Section \ref{seczoology} contains a description of a set of generic inflation models and their predictions for the form of the CMB anisotropies. Section \ref{secresults} contains results and conclusions.

\section{Inflation in scalar field theories}
\label{secinflationreview}

In this section, we quickly review scalar field models of inflationary
cosmology, and explain how we relate model parameters to observable quantities.
Inflation in its most general sense can be defined to be a period of accelerating cosmological expansion, during which the universe evolves toward homogeneity and flatness. This acceleration is typically a result of the universe being dominated by vacuum energy, with equation of state $p \simeq -\rho$. Within this broad framework, many specific models for inflation have been proposed. We limit ourselves here to models with ``normal'' gravity (i.e., general relativity) and a single order parameter for the vacuum, described by a slowly rolling scalar field $\phi$ (the {\it inflaton}). These assumptions are not overly restrictive -- most widely studied inflation models fall within this
category, including Linde's ``chaotic'' inflation scenario \cite{linde83}, inflation from pseudo Nambu-Goldstone bosons (``natural'' inflation \cite{freese90}), dilaton-like models involving exponential potentials (power-law inflation), hybrid inflation \cite{linde91,linde94,copeland94}, and so forth. Other models, such as Starobinsky's $R^2$ model\cite{starobinsky80} and versions of extended inflation, can, through a suitable transformation, be viewed in terms of equivalent single-field models.
Take a Lagrangian with a single effective degree of freedom $\phi$,
\begin{equation}
{\cal L} = {1 \over 2} g_{\mu\nu} \partial^\mu \phi \partial^\nu \phi -
V\left(\phi\right),
\end{equation}
with a metric of the flat Robertson-Walker form
\begin{equation}
ds^2 = g_{\mu\nu} dx^\mu dx^\nu = dt^2 - a^2\left(t\right) d{\bf
x}^2.\label{eqflatmetric}
\end{equation}
The {\em scale factor} $a\left(t\right)$ parameterizes the expansion of the
universe, and the expansion rate, or {\em Hubble parameter} $H$ is defined to
be
\begin{equation}
H \equiv \left({\dot a \over a}\right).
\end{equation}
For field modes homogeneous on scales comparable to the horizon size $d_H \equiv
H^{-1}$, the stress-energy of the scalar field ``matter'' is of the form of a
perfect fluid, $T_{\mu\nu} = {\rm diag}\left(\rho,-p,-p,-p\right)$, with
\begin{eqnarray}
&&\rho = {1 \over 2} \dot\phi^2 + V\left(\phi\right),\cr
&&p = {1 \over 2} \dot\phi^2 - V\left(\phi\right).
\end{eqnarray}
If the stress-energy of the universe is dominated by the scalar field, the
Einstein Field equations $G_{\mu\nu} = \left(8 \pi / M_{Pl}^2\right)
T_{\mu\nu}$ for the evolution of the metric reduce to
\begin{eqnarray}
H^2 &&= {8 \pi \over 3 M_{Pl}^2} \left[{1 \over 2} \dot\phi^2 +
V\left(\phi\right)\right],\cr
\left(\ddot a \over a\right) &&= {8 \pi \over 3 M_{Pl}^2}
\left[V\left(\phi\right) - \dot\phi^2\right].
\label{eqbackground}
\end{eqnarray}
Here $M_{Pl} = G^{-1/2} \simeq 10^{19}\ {\rm GeV}$ is the Planck mass. {\em
Inflation} is defined to be a period of accelerated expansion, $\ddot a > 0$.
The evolution of the scale factor can be given in terms of the Hubble parameter
$H$ as
\begin{equation}
a \propto \exp\left(\int{H\,dt}\right) \equiv e^N,
\end{equation}
where the number of e-folds $N$ is defined to be
\begin{equation}
N \equiv \int{H\,dt}.
\end{equation}
During inflation $H$ (and therefore the horizon size $d_H \simeq H^{-1}$) is
nearly constant, and the expansion of the universe is quasi-exponential. This
results in the curious behavior that the coordinate system is expanding faster
than the light traveling in it, and comoving length scales rapidly increase in
size relative to the horizon distance. Regions initially in causal contact are
``redshifted'' to large, non-causal scales, explaining the observed isotropy of
the cosmic microwave background (CMB) on large angular scales. This is also
important for the generation of metric fluctuations in inflation, discussed
below. Finally, a universe which starts out with a nonzero curvature evolves
rapidly during inflation toward zero curvature and a flat Robertson-Walker
metric (\ref{eqflatmetric}), which, for simplicity, we have assumed from the
beginning.

Stress-energy conservation gives the equation of motion of the scalar field,
\begin{equation}
\ddot\phi + 3 H \dot\phi + V'\left(\phi\right) = 0.
\label{eqfieldeom}
\end{equation}
The {\em slow-roll} approximation\cite{linde82,albrecht82} is the assumption
that the evolution of the field is dominated by drag from the cosmological
expansion, so that $\ddot\phi \simeq 0$ and
\begin{equation}
\dot \phi \simeq -{V' \over 3 H}.
\end{equation}
The equation of state of the scalar field is dominated by the potential,
so that $p \simeq -\rho$, and the expansion rate is approximately
\begin{equation}
H \simeq \sqrt{{8 \pi \over 3 M_{Pl}^2} V\left(\phi\right)}.
\label{eqhslowroll}
\end{equation}
The slow-roll approximation is consistent if both the slope and curvature of
the potential are small, $V',\ V'' \ll V$. This condition is conventionally
expressed in terms of the {\it slow-roll parameters} $\epsilon$ and $\eta$,
\begin{equation}
\epsilon \equiv {M_{Pl}^2 \over 4 \pi} \left({H'\left(\phi\right) \over
H\left(\phi\right)}\right)^2 \simeq {M_{Pl}^2 \over 16 \pi}
\left({V'\left(\phi\right) \over V\left(\phi\right)}\right)^2,
\end{equation}
and
\begin{equation}
\eta\left(\phi\right) \equiv {M_{Pl}^2 \over 4 \pi} \left({H''\left(\phi\right)
\over H\left(\phi\right)}\right) \simeq {M_{Pl}^2 \over 8 \pi}
\left[{V''\left(\phi\right) \over V\left(\phi\right)} - {1 \over 2}
\left({V'\left(\phi\right) \over V\left(\phi\right)}\right)^2\right].
\end{equation}
Slow-roll is then a consistent approximation for $\epsilon,\ \eta \ll 1$. The
parameter $\epsilon$ can in fact be shown to directly parameterize the equation
of state of the scalar field, $p = -\rho \left(1 - 2/3 \epsilon\right)$, so
that the second equation in (\ref{eqbackground}) can be written as
\begin{equation}
\left({\ddot a \over a}\right) = H^2 \left(1 - \epsilon\right).
\end{equation}
The condition for inflation $\ddot a > 0$ is
then simply equivalent to $\epsilon < 1$. The number of e-folds $N$ of
inflation as the field evolves from $\phi_i$ to $\phi_f$ can also be expressed
in terms of  $\epsilon$ as
\begin{equation}
N = {2 \sqrt{\pi} \over M_{Pl}} \int_{\phi_i}^{\phi_f}{d\,\phi \over
\sqrt{\epsilon\left(\phi\right)}}.
\end{equation}
To create the observed flatness and homogeneity of the universe, we
require many e-folds of inflation, typically $N \simeq 50$. This figure varies
somewhat with the details of the model. A comoving scale $k$ crosses the horizon during inflation $N\left(k\right)$ e-folds from the end of inflation, where $N\left(k\right)$ is given by\cite{lidsey97}
\begin{equation}
N(k) = 62 - \ln \frac{k}{a_0 H_0} - \ln \frac{10^{16}
        {\rm GeV}}{V_k^{1/4}}
        + \ln \frac{V_k^{1/4}} {V_e^{1/4}} - \frac{1}{3} \ln
        \frac{{V_e}^{1/4}}{\rho_{{\rm RH}}^{1/4}} \, .
\end{equation}
Here $V_k$ is the potential when the mode leaves the horizon, $V_e$ is the potential at the end of inflation, and $\rho_{{\rm RH}}$ is the energy density after reheating. Scales of order the current horizon size exited the horizon at $N\left(k\right) \sim 50 -70$. In keeping with the goal of discussing the most generic possible case, we will allow $N$ to vary within the range $50 \leq N \leq 70$ for any given model. (In the similar analysis of Ref. \cite{dodelson97}, the number of e-folds was taken to be fixed at $N = 50$.)

Inflation models not only explain the large-scale homogeneity of
the universe, but also provide a mechanism for explaining the observed level of
{\em inhomogeneity} as well. During inflation, quantum fluctuations on small scales
are quickly redshifted to scales much larger than the horizon size, where they
are ``frozen'' as perturbations in the background
metric\cite{hawking82,starobinsky82,guth82,bardeen83}. Metric perturbations at
the surface of last scattering are observable as temperature anisotropy in the
CMB, which was first detected by the Cosmic Background Explorer (COBE)
satellite\cite{smoot92,bennett96,gorski96}. The metric perturbations created during inflation are of two types:
scalar, or {\it curvature} perturbations, which couple to the stress-energy of
matter in the universe and form the ``seeds'' for structure formation, and
tensor, or gravitational wave perturbations, which do not couple to matter.
Both scalar and tensor perturbations contribute to CMB anisotropy. Scalar
fluctuations can also be interpreted as fluctuations in the density of the matter in the universe. Scalar fluctuations can be
quantitatively characterized by perturbations $P_{\cal R}$ in the intrinsic
curvature scalar\cite{mukhanov85,mukhanov88,mukhanov92,stewart93}
\begin{equation}
P_{\cal R}^{1/2}\left(k\right) = {1 \over \sqrt{\pi}} {H \over M_{Pl}
\sqrt{\epsilon}}\Biggr|_{k^{-1} = d_H}.
\end{equation}
The fluctuation power is in general a function of wavenumber $k$, and is
evaluated when a given mode crosses outside the horizon during inflation,
$k^{-1} = d_H$. Outside the horizon, modes do not evolve, so the amplitude of
the mode when it crosses back {\em inside} the horizon during a later radiation-
or matter-dominated epoch is just its value when it left the horizon during
inflation. The {\em spectral index} $n$ is defined by assuming an
approximately power-law form for $P_{\cal R}$ with
\begin{equation}
n - 1 \equiv {d\ln\left(P_{\cal R}\right) \over d\ln\left(k\right)},
\end{equation}
so that a scale-invariant spectrum, in which modes have constant amplitude at
horizon crossing, is characterized by $n = 1$. Variation of the spectral index with scale is second order in slow-roll, so we will take $n$ to be independent of scale, that is 
\begin{equation}
{d n \over d \ln k} \simeq 0.
\end{equation}
The effect of scale dependence of the spectral index is considered in Ref. \cite{copeland97}. 

Instead of specifying
the fluctuation amplitude directly as a function of $k$, it is often convenient
to specify it as a function of the number of e-folds $N$ before the end of
inflation at which a mode crossed outside the horizon. Scales of interest for
current measurements of CMB anisotropy crossed outside the horizon at $N \simeq
50\,-\,70$, so that $P_{\cal R}$ is conventionally evaluated at $P_{\cal R}\left({N =
50\,-\,70}\right)$. Similarly, the power spectrum of tensor fluctuation modes is given
by\cite{starobinsky79,rubakov82,fabbri83,abbot84,starobinsky85}
\begin{equation}
P_{T}^{1/2}\left(k_N\right) = {4 \over \sqrt{\pi}} {H \over M_{Pl}}\Biggr|_{N}.
\end{equation}
The ratio of tensor to scalar modes is then
\begin{equation}
{P_{T} \over P_{\cal R}} =  16 \epsilon,
\end{equation}
so that tensor modes are negligible for $\epsilon \ll 1$. Tensor and scalar modes both contribute to CMB temperature anisotropy. If the contribution
of tensor modes to the CMB anisotropy can be neglected, normalization to the
COBE four-year data gives\cite{bunn96,lyth96} $P_{\cal R}^{1/2} = 4.8 \times
10^{-5}$. The tensor spectral index is 
\begin{equation}
n_{T} \equiv {d \ln\left(P_{T}\right) \over d\ln\left(k\right)} = - 2 \epsilon.
\end{equation}
Note that $n_{T}$ is {\it not} an independent parameter, but is proportional to the tensor/scalar ratio,
\begin{equation}
n_{T} = - {1 \over 8} {P_{T} \over P_{\cal R}},
\end{equation}
known as the {\it consistency relation} for inflation. (This relation holds only for single-field inflation, and weakens to an inequality for inflation involving multiple degrees of freedom\cite{polarski95,bellido95,sasaki96}.) A given inflation model can therefore be described to lowest order in slow roll by three independent parameters, $P_{\cal R}$, $P_{T}$, and $n$.  
Calculating the CMB fluctuations from a particular inflationary model reduces
to the following basic steps: (1) from the potential, calculate $\epsilon$ and
$\eta$. (2) From $\epsilon$, calculate $N$ as a function of the field $\phi$.
(3) Invert $N\left(\phi\right)$ to find $\phi_N$. (4) Calculate $P_{\cal
R}$, $n$, and $P_T$ as functions of $\phi$, and evaluate at $\phi = \phi_N$, where in is in the range $N = 50\,-\,70$. For the remainder of the paper, all parameters are assumed to be evaluated at $\phi = \phi_N$.

\section{Statistics of CMB measurements: temperature and polarization}
\label{seccmbreview}

What observations of the cosmic microwave background actually measure is anisotropy in the temperature of the radiation as a function of direction. It is natural to expand the anisotropy on the sky in spherical harmonics:
\begin{equation}
{\delta T\left(\theta,\phi\right) \over T_0} = \sum_{l = 0}^{\infty}\sum_{m = -l}^{l}{a^T_{lm} Y_{lm}\left(\theta,\phi\right)},
\end{equation}
where $T_0 = 2.728^\circ K$ is the mean temperature of the CMB. Inflation predicts that each $a^T_{lm}$ will be Gaussian distributed with mean $\left\langle a^T_{lm} \right\rangle = 0$ and variance 
\begin{equation}
\left\langle a^{T*}_{l'm'} a^T_{lm}\right\rangle = C_{Tl} \delta_{ll'} \delta_{mm'}, 
\end{equation}
where angle brackets indicate an average over realizations. For Gaussian fluctuations, the set of $C_{Tl}$'s completely characterizes the temperature anisotropy. The spectrum of the $C_{Tl}$'s is in turn dependent on cosmological parameters such as $\Omega_0$, $H_0$, $\Omega_{\rm B}$ and so forth, so that observation of CMB temperature anisotropy can serve as an exquisitely precise probe of cosmological models. The parameters describing inflationary models, $P_{\cal R}$ and $P_{\cal T}$, are conventionally re-expressed in terms of quadrupole amplitudes as normalization
\begin{equation}
Q_{\rm rms-PS} \equiv T_0 \sqrt{5 C_{T2} \over 4 \pi},
\end{equation}
and tensor/scalar ratio measured at the quadrupole\cite{turner93}
\begin{equation}
r \equiv {C^{\rm tensor}_{T2} \over C^{\rm scalar}_{T2}} = 0.86 {P_T \over P_{\cal R}} = 13.7 \epsilon.
\end{equation}
Normalization is a free parameter typically determined by the self-coupling of the inflaton, so that it does not serve to constrain models. The parameters of interest for testing inflation models are the tensor/scalar ratio $r$ and the spectral index $n$. Since $r = 13.7 \epsilon$ and $n = 1 - 4 \epsilon + 2 \eta$, it is equivalent to specify the slow-roll parameters $\epsilon$ and $\eta$.

\begin{figure}
\psfig{figure=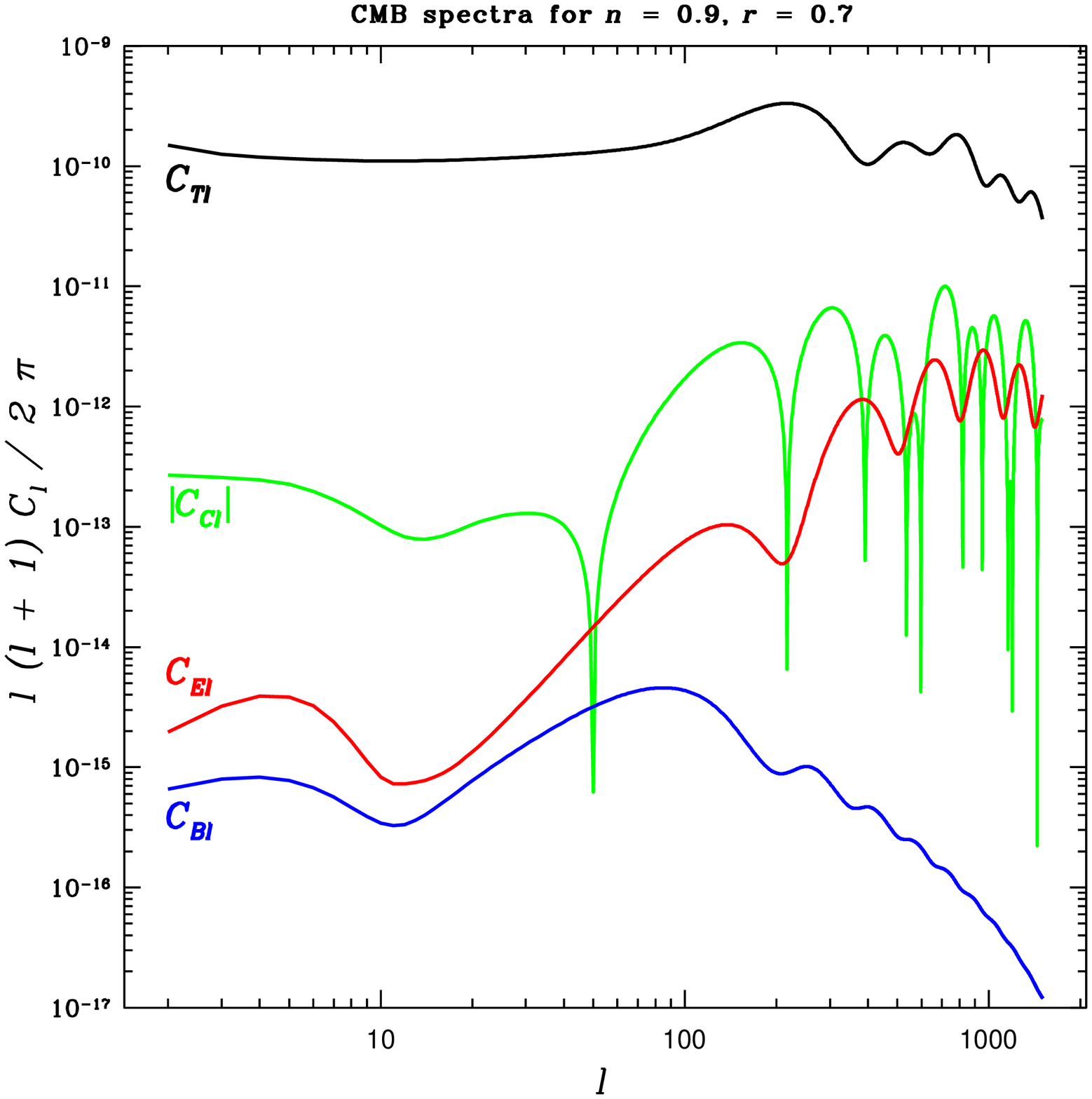,height=6.0in}
\caption{Typical temperature/polarization spectra for the fiducial case $n = 0.9$, $r = 0.7$, $\tau_{\rm ri} = 0.05$.}
\label{figspectra}
\end{figure}

Temperature anisotropy, however, is not the whole story. The cosmic microwave background is also expected to be {\it polarized} due to the presence of fluctuations. Observation of polarization in the CMB will greatly increase the amount of information available for use in constraining cosmological models. Polarization is a {\it tensor} quantity, which can be decomposed on the celestial sphere into ``electric-type'', or scalar, and ``magnetic-type'', or pseudoscalar modes. The symmetric, trace-free polarization tensor ${\cal P}_{ab}$ can be expanded as\cite{kamionkowski96}
\begin{equation}
{{\cal P}_{ab} \over T_0} =  \sum_{l = 0}^{\infty}\sum_{m = -l}^{l} \left[a^E_{lm} Y^E_{\left(l m\right) ab}\left(\theta,\phi\right) + a^B_{lm} Y^B_{\left(l m\right) ab}\left(\theta,\phi\right)\right],
\end{equation}
where the $Y^{E,B}_{\left(l m\right) a b}$ are electric- and magnetic-type tensor spherical harmonics, with parity $(-1)^l$ and $(-1)^{l + 1}$, respectively. Unlike a temperature-only map, which is described by the single multipole spectrum of $C^T_l$'s, a temperature/polarization map is described by three spectra
\begin{equation}
\left\langle \left|a^T_{lm}\right|^2\right\rangle \equiv C_{Tl},\ \left\langle\left|a^E_{lm}\right|^2\right\rangle \equiv C_{El},\ \left\langle \left|a^B_{lm}\right|^2\right\rangle \equiv C_{Bl},
\end{equation}
and three correlation functions,
\begin{equation}
\left\langle a^{T*}_{lm} a^E_{lm}\right\rangle \equiv C_{Cl},\ \left\langle a^{T*}_{lm} a^B_{lm} \right\rangle \equiv C_{(TB)l},\ \left\langle a^{E*}_{lm} a^B_{lm}\right\rangle \equiv C_{(EB)l}.
\end{equation}
Parity requires that the last two correlation functions vanish, $C_{(TB)l} = C_{(EB)l} = 0$, leaving four spectra: temperature $C_{Tl}$, E-mode $C_{El}$, B-mode $C_{Bl}$, and the cross-correlation $C_{Cl}$. Figure \ref{figspectra} shows the four spectra for a typical case. Since scalar density perturbations have no ``handedness,'' it is impossible for scalar modes to produce B-mode (pseudoscalar) polarization. Only tensor fluctuations (or foregrounds \cite{zaldarriaga98}) can produce a B-mode. 

Measurement uncertainty in cosmological parameters is characterized by the Fisher information matrix $\alpha_{ij}$. (For a review, see Ref. \cite{tegmark97}.) Given a set of parameters $\left\lbrace \lambda_i \right\rbrace$, the Fisher matrix is given by
\begin{equation}
\alpha_{ij} = \sum_l \sum_{X,Y} {\partial C_{Xl} \over \partial \lambda_i} {\rm Cov}^{-1}\left(\hat C_{Xl} \hat C_{Yl}\right)  {\partial C_{Yl} \over \partial \lambda_j},
\end{equation}
where $X,Y = T,E,B,C$ and $\rm{Cov}^{-1}\left(\hat C_{Xl} \hat C_{Yl}\right)$ is the inverse of the covariance matrix between the estimators $\hat C_{Xl}$ of the power spectra. Calculation of the Fisher matrix requires assuming a ``true'' set of parameters and numerically evaluating the $C_{Xl}$'s and their derivatives relative to that parameter choice. The covariance matrix for the parameters $\left\lbrace \lambda_i\right\rbrace$ is just the inverse of the Fisher matrix, $\left(\alpha^{-1}\right)_{ij}$, and the expected error in the parameter $\lambda_i$ is of order $\sqrt{\left(\alpha^{-1}\right)_{ii}}$. The full set of parameters $\left\lbrace \lambda_i \right\rbrace$ we allow to vary is:
\begin{enumerate}
\item{tensor/scalar ratio $r$,} 
\item{spectral index $n$,} 
\item{normalization $Q_{\rm rms-PS}$,} 
\item{baryon density $\Omega_{\rm B},$}
\item{Hubble constant $h \equiv H_0 / (100\,{\rm km\,sec^{-1}\,Mpc^{-1}})$,}
\item{reionization optical depth, $\tau_{\rm ri}$.}
\end{enumerate}
We take as a ``fiducial'' model COBE normalization \cite{bunn96} with $\Omega_{\rm B} = 0.05$ and $h = 0.5$. The results are quite sensitive to the assumed reionization history of the universe, so we separately consider the cases of no reionization ($\tau_{\rm ri} = 0$), and a reionization optical depth of $\tau_{\rm ri} = 0.05$, corresponding to reionization at a redshift of about $z \sim 13$. Note that only in the latter case do we marginalize over $\tau_{\rm ri}$. (This choice of parameters is consistent with that used by Zaldarriaga {\it et al.}\cite{zaldarriaga97a}, and, in the case of no reionization, with Dodelson {\it et al.}\cite{dodelson97}.) Fixed parameters are $\Omega_0 = 1$ and $\Omega_\Lambda = 0$, consistent with inflation.\footnote{In fact, CMB measurements are more sensitive to geometry, that is $\Omega_0 + \Omega_\Lambda$, than how the energy content of the universe is divided between matter and cosmological constant. The conclusions of this paper are essentially unchanged for the currently popular choice of $\Omega_0 = 0.3$, $\Omega_\Lambda = 0.7$. The significant assumption is a flat universe, $\Omega_0 + \Omega_\Lambda = 1$.}   For the purpose of constraining inflation, we will be interested in error ellipses projected onto the $r\,-\,n$ plane, which corresponds simply to taking the appropriate $2 \times 2$ submatrix of the full $5 \times 5$ covariance matrix $\alpha^{-1}$.
Assuming a single channel with an approximately gaussian beam, the nonzero elements of the covariance matrix ${\rm Cov}\left(\hat C_{Xl} \hat C_{Yl}\right)$ are \cite{knox95,seljak96a,zaldarriaga97b,kamionkowski96}
\begin{eqnarray}
{\rm Cov}\left(\hat C_{Tl} \hat C_{Tl}\right) =&& {2 \over \left(2 l + 1\right) f_{\rm sky}} \left(C_{Tl} + w^{-1}_T e^{l\left(l + 1\right) \sigma^2_{\rm b}}\right)^2,\cr
{\rm Cov}\left(\hat C_{El} \hat C_{El}\right) =&& {2 \over \left(2 l + 1\right) f_{\rm sky}} \left(C_{El} + w^{-1}_P e^{l\left(l + 1\right) \sigma^2_{\rm b}}\right)^2,\cr
{\rm Cov}\left(\hat C_{Bl} \hat C_{Bl}\right) =&& {2 \over \left(2 l + 1\right) f_{\rm sky}} \left(C_{Bl} + w^{-1}_P e^{l\left(l + 1\right) \sigma^2_{\rm b}} \right)^2,\cr
{\rm Cov}\left(\hat C_{Cl} \hat C_{Cl}\right) =&& {2 \over \left(2 l + 1\right) f_{\rm sky}}  \left[C^2_{Cl} + \left(C_{Tl} + w^{-1}_T e^{l\left(l + 1\right) \sigma^2_{\rm b}}\right)\left(C_{El} + w^{-1}_P e^{l\left(l + 1\right) \sigma^2_{\rm b}}\right)\right],\cr
{\rm Cov}\left(\hat C_{Tl} \hat C_{El}\right) =&& {2 \over \left(2 l + 1\right) f_{\rm sky}} C^2_{Cl},\cr
{\rm Cov}\left(\hat C_{Tl} \hat C_{Cl}\right) =&& {2 \over \left(2 l + 1\right) f_{\rm sky}} C_{Cl} \left(C_{Tl} + w^{-1}_T e^{l\left(l + 1\right) \sigma^2_{\rm b}} \right),\cr
{\rm Cov}\left(\hat C_{El} \hat C_{Cl}\right) =&& {2 \over \left(2 l + 1\right) f_{\rm sky}}  C_{Cl} \left(C_{El} + w^{-1}_P e^{l\left(l + 1\right) \sigma^2_{\rm b}} \right).
\end{eqnarray}
Here $f_{\rm sky}$ is the fraction of the sky observed, and $\sigma_{\rm b} = \theta_{\rm fwhm} / \sqrt{8 \ln 2}$ is the gaussian beamwidth, where $\theta_{\rm fwhm}$ is the full width at half maximum. The inverse weights per unit area $w^{-1}_T$ and $w^{-1}_P$ are determined by the detector resolution and sensitivity. For a noise per pixel $\sigma^T_{\rm pixel}$ and solid angle per pixel $\Omega_{\rm pixel} \simeq \theta_{\rm fwhm}^2$, the weight $w^{-1}_T$ is
\begin{equation}
w^{-1}_T = {\sigma^2_{\rm pixel} \Omega_{\rm pixel} \over T_0^2}.
\end{equation}
The polarization pixel noise $\sigma^P_{\rm pixel}$ is simply related to the temperature pixel noise $\sigma^T_{\rm pixel}$, since the number of photons available for the temperature measurement is twice that for the polarization measurements:
\begin{equation}
\left(\sigma^P_{\rm pixel}\right)^2 = 2 \left(\sigma^T_{\rm pixel}\right)^2
\end{equation}
and $w^{-1}_P = 2 w^{-1}_T$. For an observation with multiple channels $c$ with different beam sizes and sensitivities, this is easily generalized. The weights simply add, so that\cite{bond97}
\begin{equation}
{\rm Cov}\left(\hat C_{Tl} \hat C_{Tl}\right) = {2 \over \left(2 l + 1\right) f_{\rm sky}} \left(C_{Tl} + \left[\sum_{c}^{}{w^{\left(c\right)}_T e^{- l\left(l + 1\right) \left(\sigma^{\left(c\right)}_{\rm b}\right)^2}}\right]^{-1}\right)^2,
\end{equation}
and so forth. For MAP,\footnote{The specifications for MAP used here are more optimistic than the combined-channel figures of $\sigma^T_{\rm pixel} = 35\,\mu K$ and $\theta_{\rm fwhm} = 18'$ used in \cite{dodelson97}, reflecting newer figures from the MAP team. See the MAP home page at http://map.gsfc.nasa.gov.} we combine the three high-frequency channels at $40$, $60$, and $90\,{\rm GHz}$, each with a pixel noise of $\sigma_{\rm pixel} = 35\,{\rm \mu K}$ and beam sizes $\theta_{\rm fwhm} = (28.2',21.0',12.6')$ respectively. Similarly, for Planck we combine the two channels at $143$ and $217\,\rm{GHz}$, with beam width $\theta_{\rm fwhm} = (8.0',5.5')$ and  pixel noise $\sigma^T_{\rm pixel} = (5.5\,{\rm \mu K},11.7\,{\rm \mu K})$. In all cases we take the sky fraction to be $f_{\rm sky} = 0.65$.

It is particularly significant that even with a zero noise measurement, $\sigma_{\rm T} = \sigma_{\rm P} = 0$, the elements of the covariance matrix do not vanish. This means that there is an intrinsic error associated with the measurement of a given multipole $l$ of order
\begin{equation}
\left(\Delta C_{Xl} \over C_{X l}\right)^2 \simeq {{\rm Cov}\left(C_{Xl},C_{Xl}\right) \over  C_{X l}^2} \geq {2 \over \left(2 l + 1\right) f_{\rm sky}},
\end{equation}
known as {\it cosmic variance}. Cosmic variance is simply a finite sample size effect coming from the fact that we have only a single sky to measure, and is more severe at small $l$. It is in overcoming cosmic variance that precision measurement of CMB polarization holds the most dramatic promise. With temperature information only, the accuracy with which the tensor/scalar ratio $r$ can be measured is severely limited by cosmic variance, because both scalars and tensors contribute to the temperature anisotropy. With temperature information alone, $r$ of less than about $0.1$ cannot be detected, no matter how accurate the measurement. When polarization is included, arbitrarily small $r$ can in principle be detected, because only the tensor fluctuations contribute to the B-mode.

\section{A survey of inflation models}
\label{seczoology}

\begin{figure}
\psfig{figure=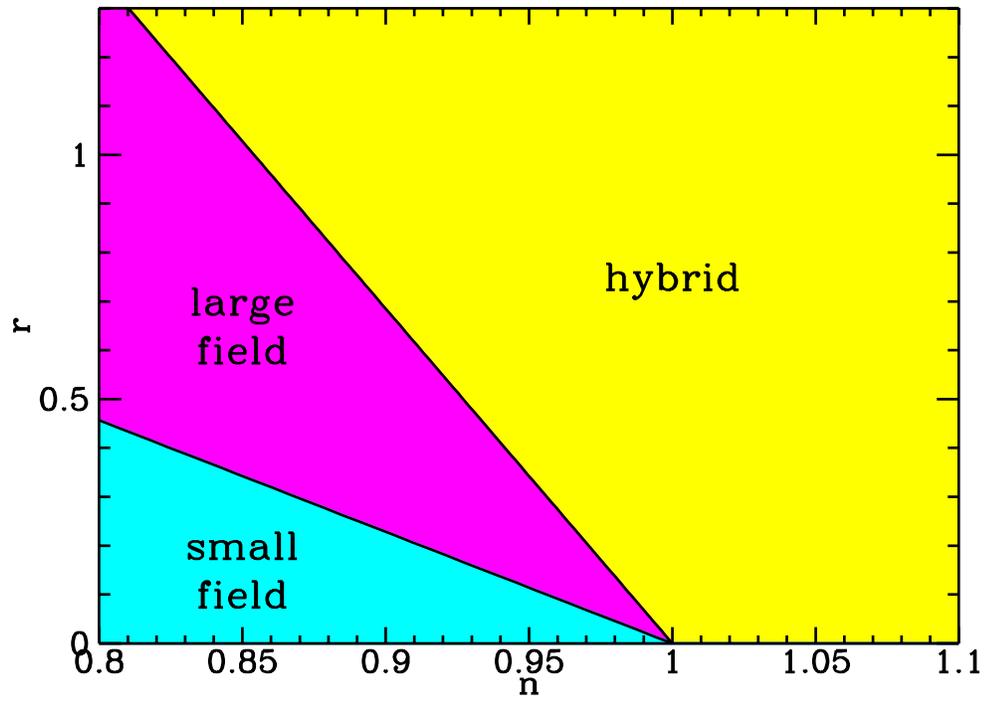,height=4.0in}
\caption{The parameter space divided into regions for small-field, large-field
and hybrid models. The linear case is the dividing line between large- and small-field.}
\label{figregions}
\end{figure}

CMB polarization can be used directly as a probe of the causal structure of the universe \cite{spergel97}. Correlations in the polarization on scales larger than the horizon at last scattering can {\it only} be produced by a period of inflationary expansion. If observed, such correlations would provide a ``smoking gun'' for inflation, regardless of any model-dependent assumptions. In this section we move from inflation in general to inflation in the particular and examine the distinct predictions of different models. Even with the restriction to single-field, slow-roll inflation, the number of models in the literature is large. It is convenient to define a general classification scheme, or ``zoology'' for models of inflation. We divide models into three general types: {\it large-field}, {\it small-field}, and {\it hybrid},  with a fourth classification, {\it linear} models, serving as a boundary between large- and small-field. A generic single-field potential  can be characterized by two independent mass scales: a ``height'' $\Lambda^4$, corresponding to the vacuum energy density during inflation, and a ``width'' $\mu$, corresponding to the change in the field value $\Delta \phi$ during inflation.  The height $\Lambda$ is fixed by normalization, so the only remaining free parameter is the width $\mu$.  Different classes of models are distinguished by the value of the second derivative of the potential, or, equivalently, by the relationship between the values of the slow-roll parameters $\epsilon$ and $\eta$\footnote{The designations ``small field'' and ``large field'' can sometimes be misleading. For instance, both Starobinsky's $R^2$ model\cite{starobinsky80} and Garc\'\i a-Bellido's ``dual inflation'' model\cite{bellido98} are characterized by $\Delta \phi \sim m_{Pl}$, but are ``small-field'' in the sense that $\eta < 0 < \epsilon$, with $n < 1$ and negligible tensor modes.}. These different classes of models have readily distinguishable consequences for the CMB. Figure \ref{figregions} shows the $r\,-\,n$ plane divided up into regions representing the large-field, small-field and hybrid cases, described in detail below.

\subsection{Large-field models: $0 < \eta \leq \epsilon$}

Large-field models are potentials typical of ``chaotic'' inflation scenarios, in which the scalar field is displaced from the minimum of the potential by an amount usually of order the Planck mass. Such models are characterized by  $V''\left(\phi\right) > 0$, and $0 < \eta \leq \epsilon$. The generic large-field potentials we consider are polynomial potentials $V\left(\phi\right) = \Lambda^4
\left({\phi / \mu}\right)^p$,
and exponential potentials, $V\left(\phi\right) = \Lambda^4 \exp\left({\phi / \mu}\right)$. 

For the case of an exponential potential, $V\left(\phi\right) \propto \exp\left({\phi / \mu}\right)$, the slow-roll parameters are constant
\begin{equation}
\epsilon = \eta = {\rm const.} = {1 \over 16 \pi} \left({M_{Pl} \over \mu}\right)^2.
\end{equation}
Models with exponential potentials are often referred to as {\it power-law} inflation, because the scale factor depends on time as a power-law, $a \propto t^{1 / \epsilon}$. The tensor/scalar ratio $r$ is simply related to the spectral index as
\begin{equation}
r = 7 \left(1 - n\right).
\end{equation}  
This result is often incorrectly generalized to all slow-roll models, but is in fact characteristic {\it only} of power-law inflation. Note that we have a one-parameter family of models, parameterized by the scale $\mu$, so that all power-law models lie on a line in the $r\,-\,n$ plane. 

For inflation with a polynomial potential $V\left(\phi\right) \propto \phi^p$, the slow-roll parameters are
\begin{equation}
\epsilon = {p \over p + 4 N} = \left({p \over p - 2}\right) \eta,\label{eqlfetaoverep}
\end{equation}
where N is the number of e-folds. Again we have $r \propto 1 - n$, 
\begin{equation}
r \simeq 7 \left({p \over p + 2}\right) \left(1 - n\right).
\end{equation}
so that tensor modes are large for significantly tilted spectra. Unlike the case of the exponential potential, the scale $\mu$ drops out of the expressions for the observables, and the models are parameterized by the {\it discrete} exponent $p$.

\subsection{Small-field models: $\eta < 0 < \epsilon$}

Small-field models are the type of potentials that arise naturally from spontaneous symmetry breaking. The field starts from near an unstable equilibrium (taken to be at the origin) and rolls down the potential to a stable minimum. Small-field models are characterized by $V''\left(\phi\right) < 0$ and $\eta < 0 < \epsilon$. The generic small-field potentials we consider are of the form $V\left(\phi\right) = \Lambda^4 \left[1 - \left({\phi / \mu}\right)^p\right]$, which can be viewed as a lowest-order Taylor expansion of an arbitrary potential about the origin\cite{kinney96a,kinney96b}. Assuming $\left(\phi_N / \mu\right) \ll 1$, the slow-roll parameters can be related as
\begin{equation}
\epsilon = {p \over 2 \left(p - 1\right)} \left| \eta \right| \left({\phi_N \over \mu}\right)^p,\label{eqsfepsilon}
\end{equation}
where
\begin{equation}
\eta \simeq - {p \left(p - 1\right) \over 8 \pi} \left({M_{Pl} \over \mu}\right)^2 \left({\phi_N \over \mu}\right)^{p - 2}.\label{eqetasmallfield}
\end{equation}
In general, then, $\epsilon \ll \left|\eta\right|$ and the spectral index is approximately
\begin{equation}
n \simeq 1 + 2 \eta.
\end{equation}
This leads to a simple relationship between the tensor/scalar ratio and the spectral index
\begin{equation}
r = 7 \left(1 - n\right) \left[{p \over 2 \left(p - 1\right)} \left({\phi_N \over \mu}\right)^p\right],
\end{equation}
so that tensor modes are strongly suppressed in small-field models relative to the large-field case. Because of the exponent $p - 2$ in the expression for $\eta$ (\ref{eqetasmallfield}), the cases $p = 2$ and $p > 2$ have very different behavior. For $p = 2$, taking inflation to end at $\phi_{\rm E} \sim \mu$, it is straightforward to compute $\phi_N$:
\begin{equation}
\left({\phi_N \over \mu}\right) \simeq \exp\left[- {1 \over 2} - {1 \over 2} N \left(1 - n\right)\right].
\end{equation}
We then have the desired expression for $r$ as a function of $n$:
\begin{equation}
r = 7 (1 - n) \exp\left[- 1 - N\left(1 - n\right)\right].
\end{equation}
For $p > 2$, the field value $\phi_N$ is
\begin{equation}
\left({\phi_N \over \mu}\right)^{p - 2} = {8 \pi \over N p \left(p - 2\right)} \left({\mu \over M_{Pl}}\right)^2,
\end{equation}
so that
\begin{equation}
\eta \simeq  - {1 \over N} \left({p - 1 \over p - 2}\right).
\end{equation}
The scalar spectral index is then
\begin{equation}
n \simeq 1 - {2 \over N} \left({p - 1 \over p - 2}\right),
\end{equation}
which is {\it independent} of $\left(\mu / M_{Pl}\right)$, so that $\Delta \phi \sim \mu \ll M_{Pl}$ is consistent with a nearly scale-invariant scalar fluctuation spectrum\cite{kinney96a,kinney96b}. We can make $\left(\mu / M_{Pl}\right)$ as small as we wish, so that the tensor/scalar ratio
\begin{equation}
r \propto \left({\mu \over M_{Pl}}\right)^{2 p / \left(p - 2\right)}
\end{equation}
can be arbitrarily small for $p > 2$. In keeping with the physical motivation for these models, we take $\mu < M_{Pl}$, so that there is an upper bound on $r$ of
\begin{equation}
r < 7 {p \over N \left(p - 2\right)} \left({8 \pi \over N p \left(p - 2\right)}\right)^{p / \left(p - 2\right)}.
\end{equation}
It is particularly interesting to consider the question of whether measurement of the polarization will make it possible to distinguish between $p = 2$ and $p > 2$ small-field models. This is discussed in Section \ref{secresults}.

\subsection{Hybrid models: $0 < \epsilon < \eta$}

The hybrid scenario frequently appears in models which incorporate inflation into supersymmetry. In a hybrid inflation scenario, the scalar field responsible
for inflation evolves toward a minimum with nonzero vacuum energy. The end of inflation arises as a result of instability in a second field. Hybrid models are characterized by $V''\left(\phi\right) > 0$ and $0 < \epsilon < \eta$. We consider generic potentials for hybrid inflation of the form $V\left(\phi\right) =
\Lambda^4 \left[1 + \left({\phi / \mu}\right)^p\right].$ The field value at the end of inflation (and hence $\phi_{N}$) is determined by some other physics, and we treat $\left(\phi_{N} / \mu\right)$ in this case as a freely adjustable parameter. The slow-roll parameters are then related as
\begin{eqnarray}
\frac{\eta }{\epsilon }& = &
\frac{2 \left( p - 1\right)}{p} \left(\frac{\phi_{N}}{\mu}\right)^{-p}
\left[1 + \frac{p - 2}{2 \left(p - 1\right)}
        \left(\frac{\phi_{N}}{\mu}\right)^p\right]  \cr
 & \longrightarrow & \left\{ \begin{array}{ll}
\frac{{\displaystyle p-2}}{{\displaystyle p}} & {\mbox {for\ }}\phi_{N}/\mu \gg 1 \\  & \\
\frac{{\displaystyle 2 \left( p - 1\right)}}{{\displaystyle p}}
\left( \frac{{\displaystyle \mu}}{{\displaystyle \phi_{N}}}\right)^p &
                {\mbox {for\ }}    \phi_{N}/\mu \ll 1
\end{array} \right.   .
\end{eqnarray}
When $\left(\phi_N / \mu\right) \gg 1$, we recover the result for the large-field case (\ref{eqlfetaoverep}). When $\left(\phi_N / \mu\right) \ll 1$, we obtain a result analogous to that for small-field models (\ref{eqsfepsilon}), with the difference that here $\eta$ is positive.  The distinguishing feature of many hybrid models is a {\it blue} scalar spectral index, $n > 1$. This corresponds to the case $\eta > 2 \epsilon$. Recalling that $n = 1 - 4 \epsilon + 2 \eta$, we see that hybrid models can also in principle have a {\em red} spectrum, $n < 1$. Because of the extra freedom to choose the field value at the end of inflation, and hence $\left(\phi_N / \mu\right)$, hybrid models fill a broad region in the $r\,-\,n$ plane (Fig. \ref{figregions}). There is, however, no overlap in the $r\,-\,n$ plane between hybrid inflation and other models.

\subsection{Linear models: $\eta = - \epsilon$}

Linear models, $V\left(\phi\right) \propto \phi$, live on the boundary between
large-field and small-field models, with $V''\left(\phi\right) = 0$ and $\eta = - \epsilon$. We then have the relation 
\begin{equation}
r = {7 \over 3} \left(1 - n\right).
\end{equation}

This enumeration of models is certainly not exhaustive. There are a number of single-field models that do not fit well into this scheme, for example logarithmic potentials\cite{barrow95} $V\left(\phi\right) \propto \ln\left(\phi\right)$ typical of sypersymmetry. Another example is potentials with negative powers of the scalar field $V\left(\phi\right) \propto \phi^{-p}$ used in intermediate inflation \cite{barrow93} and dynamical supersymmetric inflation \cite{kinney97,kinney98}. However, the three classes categorized by the relationship between the slow-roll parameters as $0 < \eta \leq \epsilon$ (large-field), $\eta < 0 < \epsilon$ (small-field), and $0 < \epsilon < \eta$ (hybrid), cover the entire $r\,-\,n$ plane and are in that sense complete. 

\section{Results and Conclusions}
\label{secresults}

\begin{figure}
\psfig{figure=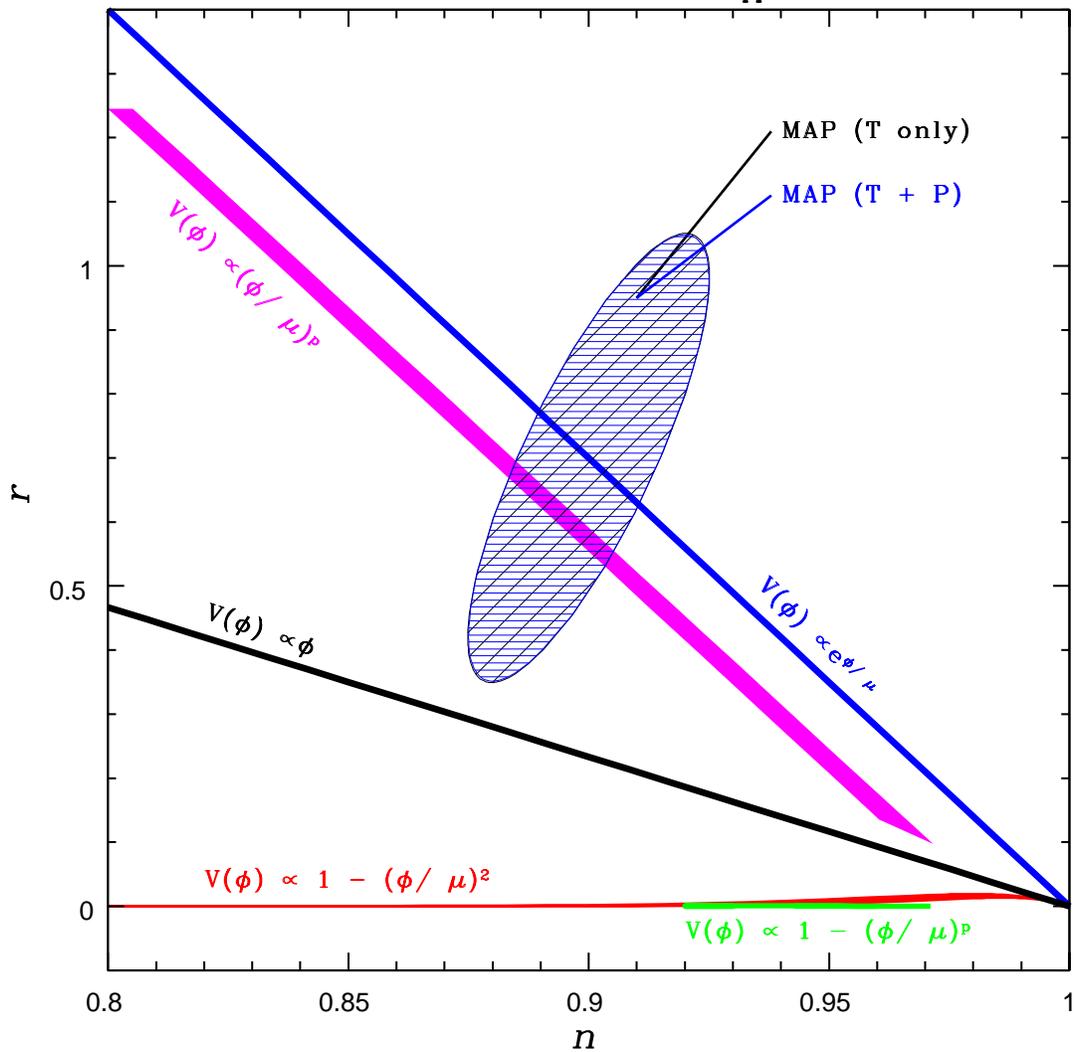,height=6.0in}
\caption{$2-\sigma$ error ellipses for MAP (no reionization), with $n = 0.9$ and $r = 0.7$. The T only and T+P ellipses overlap almost exactly in this case.}
\label{figmapnoreio}
\end{figure}

\begin{figure}
\psfig{figure=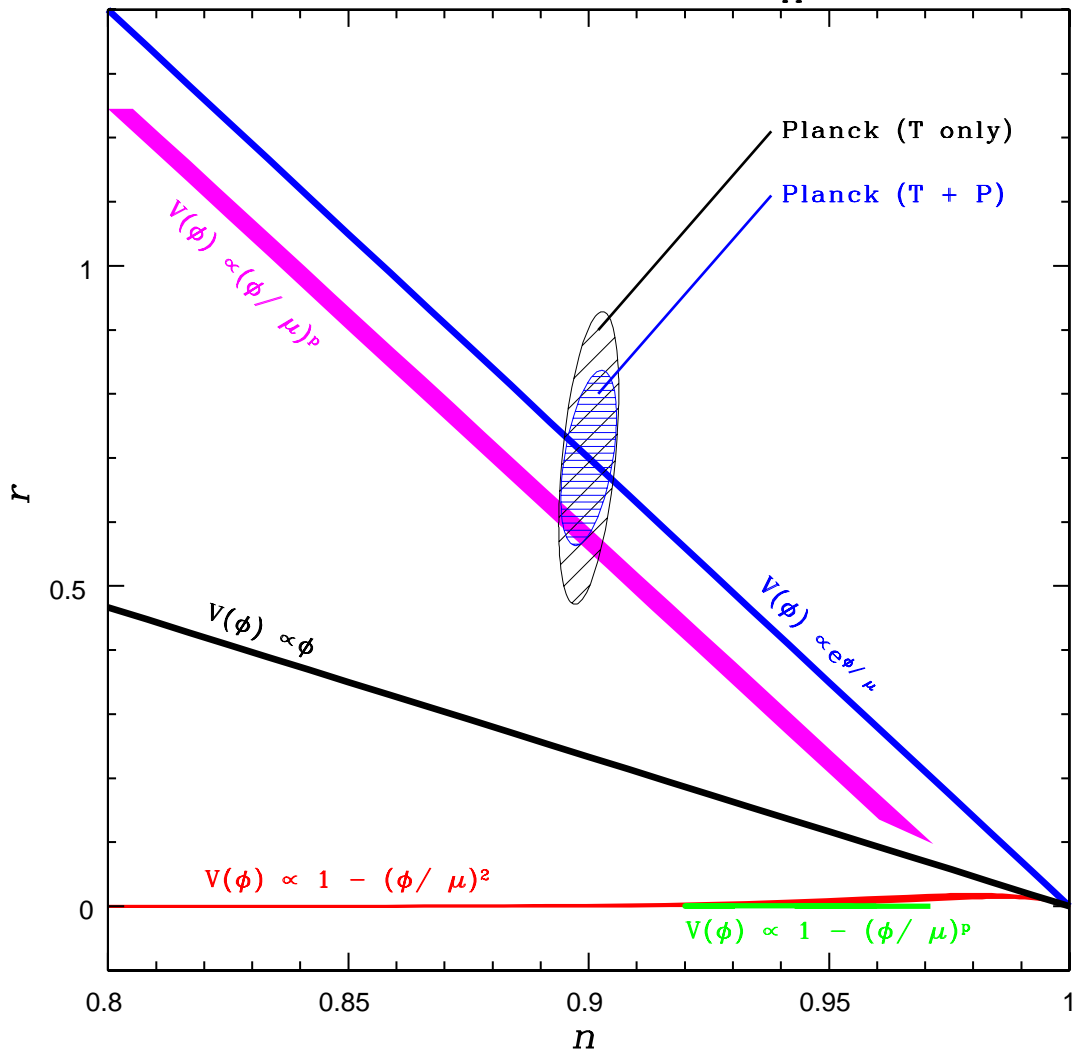,height=6.0in}
\caption{Error ellipses for Planck (no reionization), with $n = 0.9$ and $r = 0.7$. In this case, polarization results in a noticeable increase in sensitivity.}
\label{figplancknoreio}
\end{figure}

\begin{figure}
\psfig{figure=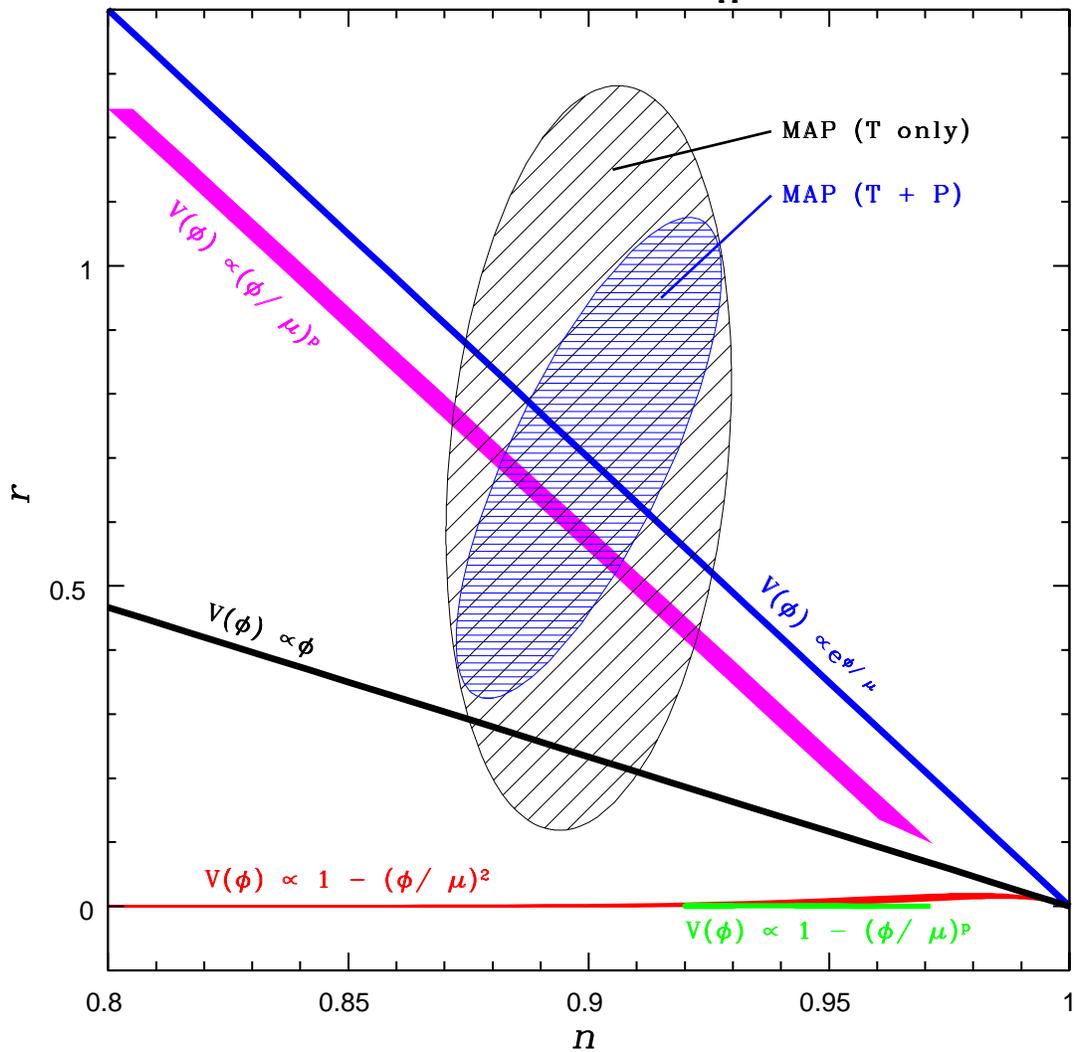,height=6.0in}
\caption{Error ellipses for MAP ($\tau_{\rm ri} = 0.05$). With reionization, the T-only error is much larger than for the case with no reionization, but the T+P case is mostly unaffected.}
\label{figmaplss}
\end{figure}

\begin{figure}
\psfig{figure=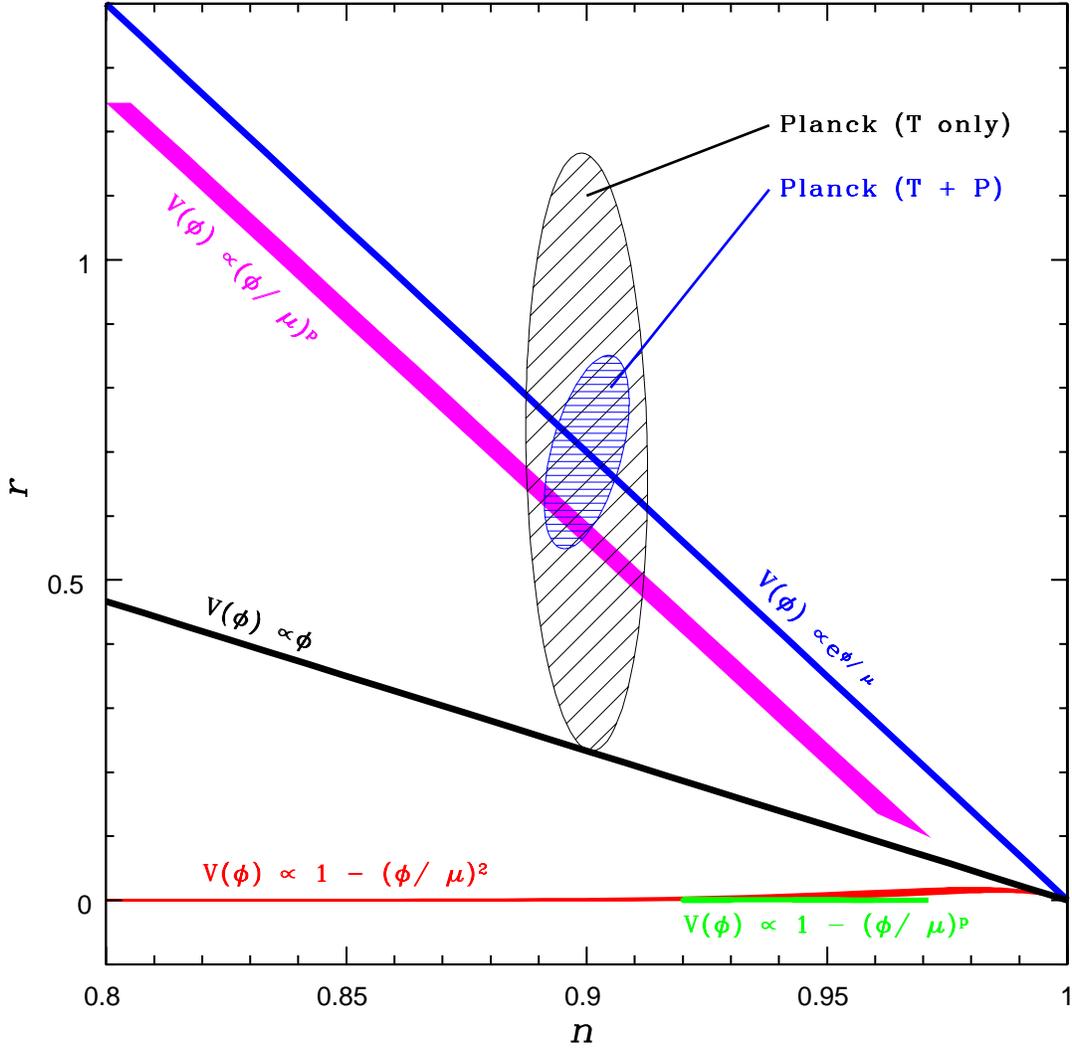,height=6.0in}
\caption{Error ellipses for Planck ($\tau_{\rm ri} = 0.05$). As with MAP, the temperature-only measurement is degraded by reionization, but not the errors including polarization information.}
\label{figplancklss}
\end{figure}

\begin{figure}
\psfig{figure=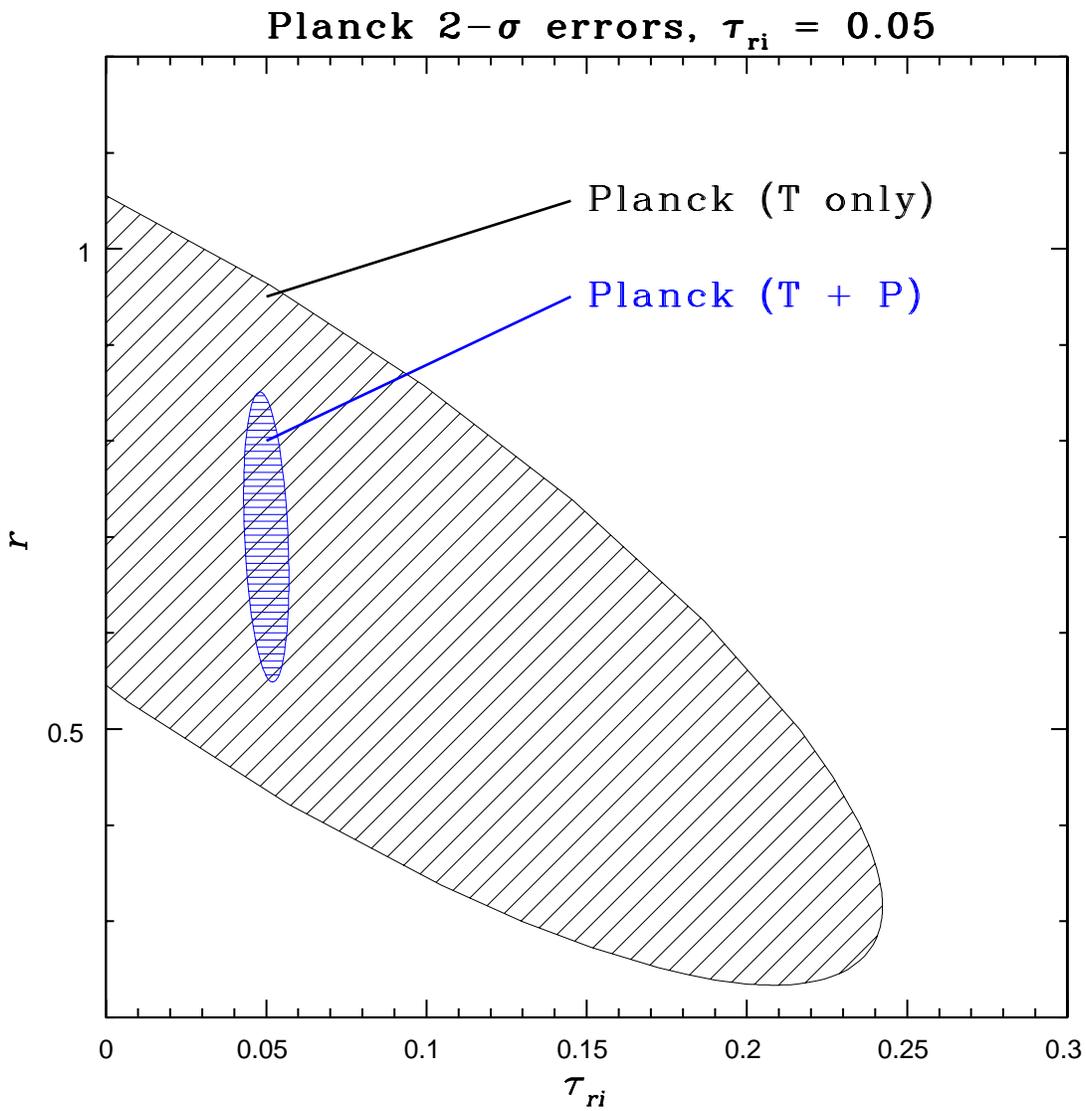,height=6.0in}
\caption{Error ellipses in the $\tau_{\rm ri}\,-\,r$ plane. The parameter degeneracy present in the T-only measurement is removed completely by including polarization information, a significant advantage.}
\label{figtaur}
\end{figure}

The goal is to answer two questions: first, for a model with easily detectable tensor modes, how much does measurement of CMB polarization increase the precision of the measurement in the $r\,-\,n$ parameter plane? Second, what is the smallest $r$ detectable under a reasonable set of assumptions? The ability to detect an $r$ of about $0.01$ would make it possible to observationally distinguish between small-field models with quadratic potentials and models with higher powers of $\phi$.
We calculate the $C_l$ spectra using Seljak and Zaldarriaga's CMBFAST code \cite{seljak96b}, and plot the expected errors for MAP and Planck on the $r\,-\,n$ plane, along with the predictions of the generic inflation models discussed in Section \ref{seczoology}.  The size of the expected errors depend, of course, on the underlying model assumed. In all cases we take COBE normalization, $h = 0.5$, and $\Omega_b = 0.05$. We choose two models for study. The first model, motivated by consideration of large-scale structure data \cite{white95}, is on the power-law inflation curve with $n = 0.9$ and $r = 0.7$. Figure \ref{figmapnoreio} shows MAP $2-\sigma$ error ellipses with and without polarization in the $r - n$ plane, assuming no reionization. Note in particular that the error ellipses with and without polarization information are identical in size -- observing polarization does not help in this case. Figure \ref{figplancknoreio} shows the equivalent result for Planck. In this case, polarization increases the precision of the measurement. Figures \ref{figmaplss} and \ref{figplancklss} show the errors for MAP and Planck with an assumed reionization optical depth of $\tau_{\rm ri} = 0.05$, corresponding to a reionization redshift $z_{\rm ri} \simeq 13$. In this case the errors on the temperature-only measurement are much larger than in the case where we assumed no reionization, but the errors when polarization information is included are practically unchanged. The conclusion to be drawn is that measuring CMB polarization eliminates the degeneracy between reionization and tensor modes present in temperature-only measurements, a significant advantage. This can be seen more clearly by plotting the error ellipse in the $\tau_{\rm ri}\,-\,r$ parameter plane (Fig. \ref{figtaur}). With a temperature-only measurement, there is significant degeneracy between $\tau_{\rm ri}$ and $r$, but the degeneracy disappears completely when polarization is included. 

\begin{figure}
\psfig{figure=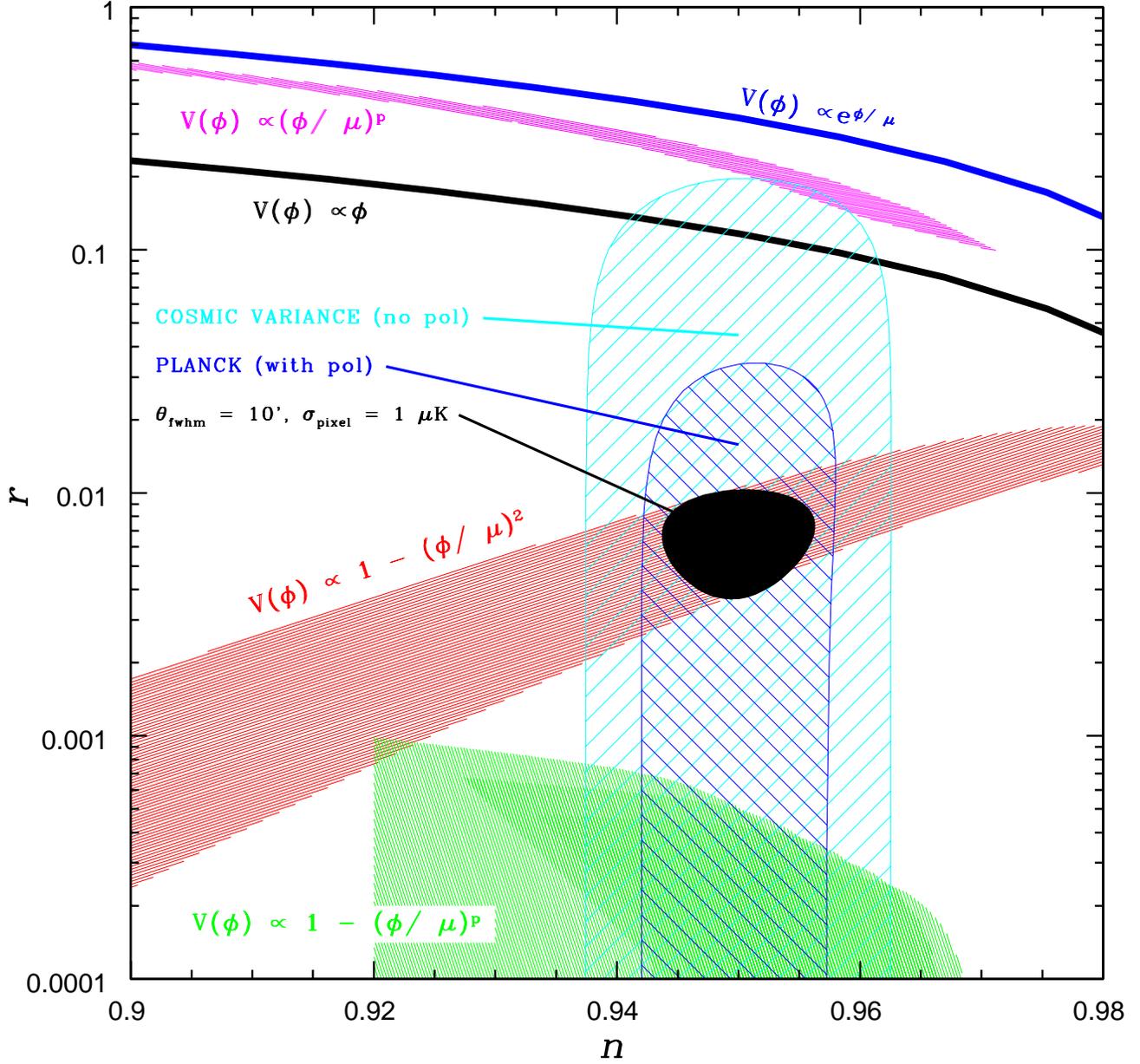,height=7.0in}
\caption{The $r\,-\,n$ plane on a logarithmic scale, highlighting the predictions of small-field models. Error ellipses are for cosmic variance ($\sigma^T_{\rm pixel} = 0$), Planck (with polarization), and a hypothetical experiment with the same $10'$ angular resolution as Planck but a factor of three higher sensitivity. This is sufficient to detect $r \sim 0.01$, which makes is possible to distinguish between $p = 2$ and $p > 2$ small-field models.}
\label{figsmallr}
\end{figure}

The second model is in the region predicted by a small-field model with $V\left(\phi\right) \propto 1 - \left(\phi / \mu\right)^2$, with $r = 0.007$ and $n = 0.95$. Figure \ref{figsmallr} again shows the inflation models plotted on the $r\,-\,n$ plane, this time with $r$ on a logarithmic scale.  Error ellipses are shown for the cosmic variance limit of a temperature-only measurement (that is, for a {\it perfect} temperature-only observation, assuming a resolution of $10'$), for Planck (with polarization), and for a hypothetical all-sky measurement with a $10'$ resolution (similar to Planck), but with  $\sigma_{\rm pixel} = 1\mu K$, roughly a factor of three higher than the figure for Planck's combined-channel sensitivity. The reionization optical depth is assumed to be $\tau_{\rm ri}$ = 0.05. We see that while Planck with polarization does better than cosmic variance, it still is not sensitive enough to detect the tensor mode in this case. However, a small improvement in sensitivity gives a big payoff: the ability to distinguish experimentally between $p = 2$ and $p > 2$ small-field inflation models, something which is impossible using temperature information alone.  These results are in good agreement with the conclusions of Kamionkowski and Kosowsky \cite{kamionkowski97}.

While models of inflation make the generic predictions of a flat universe and nearly scale-invariant spectrum of scalar fluctuations, different models make quite distinct predictions within that general framework. A previous paper \cite{dodelson97} considered measurements of the CMB temperature fluctuations to show that MAP and Planck would be capable of discriminating, at least roughly, between different classes of inflation models. While such information can't ``prove'' a model correct, it can make it possible to eliminate models as inconsistent with observation. In this paper, we have extended that analysis to include CMB polarization  as well as temperature, with two main conclusions. First, reionization can significantly degrade the accuracy of temperature-only measurements in the parameter space of relevance for discriminating among inflation models. However, the inclusion of polarization largely eliminates this problem.  Second, very small tensor to scalar ratios can be probed with polarization, given that there is no fundamental lower limit from cosmic variance. While planned satellite experiments are not capable of detecting a small enough $r$ to distinguish between small-field models with a quadratic potential and those dominated by higher powers of $\phi$, a factor of three improvement in sensitivity over Planck would make such discrimination possible. It is also conceivable that ground-based experiments might trade decreased sky coverage for increased sensitivity with the same overall result. CMB polarization opens the door to precision tests of inflation models.

\section*{Acknowledgments}

This work has benefited greatly from conversations with Edward Kolb, Marc Kamionkowski, Scott Dodelson, Albert Stebbins, Gary Hinshaw, Mathias Zaldarriaga, and Lloyd Knox.

This work was supported in part by DOE and NASA grant NAG5-7092 at Fermilab.

\end{document}